\documentclass[draft]{agujournal2019}
\usepackage{url} 
\usepackage{lineno}
\usepackage[finalnew]{trackchanges} 
\usepackage{soul}
\newcommand{\dd}{\mathrm{d}}
\usepackage{bm}
\usepackage{siunitx}
\draftfalse

\journalname{JGR: Space Physics}

\begin{document}

\title{Laboratory verification of electron-scale reconnection regions modulated by a three-dimensional instability}

\authors{S. Greess\affil{1}, J. Egedal\affil{1}, A. Stanier\affil{2}, W. Daughton\affil{2}, J. Olson\affil{1}, A. L\^{e}\affil{2}, R. Myers\affil{1}, A. Millet-Ayala\affil{1}, M. Clark\affil{1}, J. Wallace\affil{1}, D. Endrizzi\affil{1}, C. Forest\affil{1},}

\affiliation{1}{University of Wisconsin - Madison}
\affiliation{2}{Los Alamos National Laboratory, Los Alamos, New Mexico 87545, USA}

\correspondingauthor{Samuel Greess}{greess@wisc.edu}

\begin{keypoints}

\item The structure of the electron diffusion region is important to reconnection in space plasma
\item The widths of laboratory electron reconnection layers match those of kinetic simulations
\item Kinetic simulations show that the electron pressure tensor breaks the electron frozen flux condition
\end{keypoints}

\begin{abstract}
During magnetic reconnection in collisionless space plasma, the electron fluid decouples from the magnetic field within narrow current layers, and theoretical models for this process can be distinguished in terms of their predicted current layer widths. From theory, the off-diagonal stress in the electron pressure tensor is related to thermal non-circular orbit motion of electrons around the magnetic field lines. This stress becomes significant when the width of the reconnecting current layer approaches the small characteristic length scale of the electron motion. To aid {\sl in situ} spacecraft and numerical investigations of reconnection, the structure of the electron diffusion region is here investigated using the Terrestrial Reconnection EXperiment (TREX). In agreement with the closely matched kinetic simulations, laboratory observations reveal the presence of electron-scale current layer widths. Although the layers are modulated by a current-driven instability, 3D simulations demonstrate that it is the off-diagonal stress that is responsible for breaking the frozen-in condition of the electron fluid. 
\end{abstract}

\section*{Plain Language Summary}
``Space weather'' describes the conditions of the plasma surrounding Earth, which can have severe impact on the functionality of spacecraft as well as the health of human space travelers. Space weather and the dynamics of space plasmas in general are closely linked to the structure and topology of the magnetic fields permeating our solar system. By a process called magnetic reconnection, magnetic field lines can rapidly and suddenly break and reconfigure their connectivity, allowing for an explosive release of magnetic energy. This phenomenon is at the origin of explosive events such as solar flares and is the driver of magnetic storms in the Earth's magnetosphere powering the Auroras. 

We present new laboratory observations of this near-Earth reconnection process recreated in the Terrestrial Reconnection EXperiment (TREX). The experiment provides detailed measurements of the width of the region where the magnetic field lines break, the electron diffusion region (EDR). Consistent with supercomputer simulations of reconnection, the width of the EDR is measured to be set by the fine spatial scale of the electron orbit motions. As such, the observations provide renewed evidence that the reconnection process is mediated by forces of thermal stress related to the electron motion within the reconnection region.

\section[sec:intro]{Introduction}

Magnetic reconnection \cite{dungey:1953} is the process of changing the topology of magnetic field lines in the presence of a plasma, often permitting an explosive release of magnetic energy. Well-known examples include solar flares \cite{priest:2000} and auroral substorms in the Earth's magnetosphere \cite{vasyliunas:1975}. Although reconnection often governs the global dynamics of plasma systems, the reconnection process occurs in localized electron diffusion regions (EDRs), where the motion of the electron fluid decouples from the magnetic field, breaking the frozen-in law of magnetohydrodynamics. The origin of this process in the collisionless regime, where conventional resistive friction is absent, remains controversial. For example, laminar kinetic models predict that the EDRs are characterized by intense current layers with widths as narrow as the kinetic scales associated with the electron orbit motion \cite{vasyliunas:1975,pritchett:2001}. In other models, the scattering of electrons by electric field fluctuations associated with high-frequency instabilities is proposed to widen the current layers and enhance the anomalous transfer of momentum from the electrons to the ions \cite{papadopoulos:1977,huba:1977,hoshino:1991}.

Significant insight into reconnection physics is provided by fully kinetic numerical models. In 3D configurations it has been argued that turbulence can cause local suppression of the effective conductivity \cite{silin:2005,che:2011,munoz:2017}, but other simulation studies have reported these effects are relatively small in both low-$\beta$ parameter regimes \cite{liu:2013} relevant to solar physics and for higher-$\beta$ regimes with asymmetric layers relevant to the magnetosphere \cite{roytershteyn:2012,hesse:2018,le:2018}. Rather, these 2D and 3D kinetic models typically suggest that fast reconnection can be mediated by electron inertia, and terms in the electron pressure tensor
\cite{speiser:1965,vasyliunas:1975,lyons:1990,horiuchi1994particle,cai1997generalized,kuznetsova1998kinetic}. These effects require the formation of intense electron current channels with widths characterized by either the electron inertial length $d_e=c/\omega_{pe}$ or the electron orbit scale \cite{roytershteyn:2013}. 

To observationally address this issue, a primary goal of NASA's Magnetospheric Multiscale (MMS) Mission is to characterize the structures of EDRs for reconnection sites in the Earth's magnetosphere \cite{burch:2016b}. However, the {\sl in situ} observations have not as yet provided conclusive insight to the role of anomalous resistivity. For example, the initial magnetotail observations are consistent with laminar kinetic reconnection \cite{torbert:2018,egedal:2019,genestreti2018,nakamura2018}. Meanwhile, for a magnetopause reconnection layer crossing \cite{burch:2016} evidence for anomalous resistivity was identified near an EDR \cite{torbert:2016}, but a separate analysis concluded the electron dynamics were in agreement with a 2D kinetic model (without anomalous resistivity) \cite{egedal:2018}. 

Dedicated laboratory experiments can provide complementary methods to study EDRs. Contrary to spacecraft measurements, laboratory experiments allow the controlled and reproducible study of reconnection layers with well understood upstream conditions and magnetic geometry. Results from the Magnetic Reconnection Experiment (MRX) at Princeton find that the current layer widths are much wider (by approximately a factor of four) compared to the predictions by kinetic models \cite{ji:2008}. This disagreement \cite{dorfman:2008} remains unresolved as it persists even when accounting for collisions \cite{roytershteyn:2010}, and 3D instabilities \cite{roytershteyn:2013}.

In this paper, we report on experimental and numerical investigations of the EDR width using the Terrestrial Reconnection EXperiment (TREX) at the University of Wisconsin-Madison \cite{Forest2015,olson:2016,olson:2021}. While TREX and MRX both have similar normalized system sizes, $\simeq 10 c/\omega_{pi}$ where $\omega_{pi}$ is the ion plasma frequency, TREX is physically larger and operates at lower plasma density and a more collisionless regime (Lundquist number, $S\simeq 10^4$), where anisotropic and non-gyrotropic electron pressure tensor effects can begin to develop around the EDR \cite{le:2015}. Furthermore, the TREX experiment applies a unique jogging method, where the reconnection layer is swept across magnetic sensors, yielding high spatial resolution measurements of the magnetic structures. 
To elucidate the experimental findings, fully kinetic 2D and 3D simulations were performed with plasma profiles and reconnection drives that are comparable to the experiment.

\section[sec:trex]{The Terrestrial Reconnection EXperiment (TREX)}

\begin{figure}
\includegraphics[width=\textwidth]{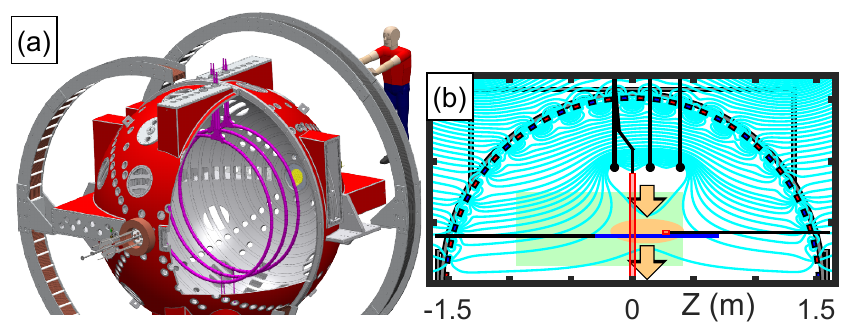}
\caption{(a) Engineering sketch of TREX. The internal drive coils (purple) drive a magnetic field that opposes the external Helmholtz coil's field. The plasma source is a polar array of plasma guns (yellow). (b) A cross-section of the top half of the TREX vessel showing a theoretical example of the typical experimental geometry. The magnetic field lines are shown in cyan. The reconnection region (light orange) is driven down from the drive coils to the central axis, as indicated by the arrows. The layer is measured during this transit by the three probes shown: the 3-axis linear $\mathbf{\dot{B}}$ probe array (blue), the speed probe (long red), and the multi-tip Langmuir temperature/density probe, known as the $T_e$ probe (short red). The hook probe, another array of 3-axis $\mathbf{\dot{B}}$ probes can be scanned through the shaded green area, allowing for the compilation of data from multiple experimental shots. These probes operate on sampling frequencies on the order of $10\si{\mega\hertz}$, while the elapsed time between the layer's generation and its arrival at the central axis is on the order of $20\si{\micro\second}$}.
\label{fig:schematic}
\end{figure}

The applied TREX configuration is presented by the engineering schematic in Fig.~\ref{fig:schematic}(a). The vacuum vessel, provided by the Wisconsin Plasma Physics Laboratory (WiPPL) \cite{Forest2015}, is a $3$ meter diameter sphere that uses an array of permanent magnets embedded in the chamber wall to limit the plasma loss area to a very narrow fraction of the total surface area while keeping the bulk of the plasma unmagnetized. The setup includes a set of internal drive coils and an exterior Helmholtz coil that provides a near-uniform axial magnetic field with a magnitude up to $100\si{\milli\tesla}$ \cite{olson:2016,olson:2021}. The current through the three internal drive coils (purple) ramps up to create a magnetic field that opposes and reconnects with the background Helmholtz field, resulting in an anti-parallel magnetic configuration (e.g. no significant guide field). The plasma source is a set of plasma guns located at the machine's pole (shown in yellow). TREX can operate in hydrogen, deuterium and helium plasmas; results presented in this paper will focus on hydrogen and deuterium.\add[]{ As will be described in the following sections, the interpretation of the experimental results are aided by 2D simulations at the full hydrogen/electron mass ratio ($m_i/m_e=1836$). For numerical tractability, 3D simulations were implemented with a mass ratio of $m_i/m_e=400$.}

In the planar cut of TREX shown in Fig.~\ref{fig:schematic}(b), the cyan lines are theoretically-derived magnetic flux contours meant to illustrate the typical magnetic geometry of an experimental run. As the current through the drive coils ramps up, the reconnection region is pushed from underneath the drive coils radially inward (orange arrows in Fig.~\ref{fig:schematic}(b)). During this transit, the reconnection layer ``jogs" past the electrostatic and magnetic probes. Given the near constant speed of the reconnection layer, this facilitates high spatial resolution measurements of the entire layer geometry over the course of a single experimental shot; this type of measurement is referred to as the jogging method. These probes and their locations are represented by the blue and red rectangles in Fig.~\ref{fig:schematic}(b). In addition to these jogging method probes, a different array of 3-axis $\mathbf{\dot{B}}$ probes can be moved between shots, allowing for the creation of multi-shot datasets. The coverage area of this probe is given by the light green rectangle in Fig.~\ref{fig:schematic}(b). By compiling data from multiple shots taken at different location, this probe provides information about the reconnection geometry without relying on the jogging method.

\begin{figure}
\includegraphics[width=\textwidth]{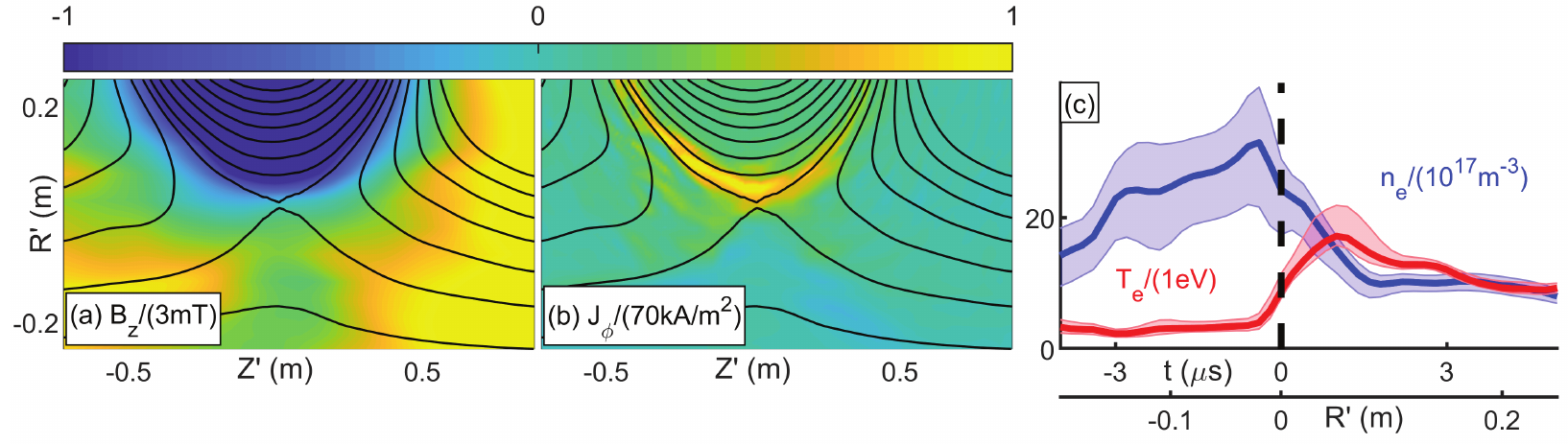}
\caption{Example of experimental data. Plots (a) and (b) show the data from the hook probe recorded in a scan including $34$ different probe positions covering the green region in Fig.~\ref{fig:schematic}(b). The black lines are contours of the flux function, $\Psi$, which map to the magnetic field lines. (a) shows the reconnecting magnetic fields, and (b) shows the out-of-plane current layer. (c) shows data from the $T_e$ probe (short red in Fig.~\ref{fig:schematic}(b)); the shaded regions represent the $95\%$ confidence interval for the values of density and temperature based on the fit of the probe's IV curve. The lower $R$ side of the layer is closer to the plasma sources and thus has a higher density than the other side of the layer. There is a jump in the plasma temperature when the layer passes the probe. The data in (c) is compiled using the jogging method to convert the time signal into a measurement of the $R$-coordinate.}
\label{fig:exampledata}
\end{figure}

An example of data collected from a typical set of experimental shots is provided in Fig.~\ref{fig:exampledata}, where Fig.~\ref{fig:exampledata}(a-b) shows data from $34$ shots combined into one picture; for each shot, the hook probe is at a different position within the green region in Fig.~\ref{fig:schematic}(b). The black lines are contours of the flux function $\Psi$ to illustrate the in-plane magnetic field lines. Fig.~\ref{fig:exampledata}(c) shows the temperature and density data measured by the $T_e$ probe. Note that in Fig.~\ref{fig:exampledata}(c), the time signals from the $T_e$ probe are converted into position data using the jogging method described above. Typical plasma parameters include $T_i\ll T_e\simeq 5 - 20 \si{\electronvolt}$, $n_e\simeq 2\cdot10^{18} \si{\meter}^{-3}$, $B_{rec}\simeq 4\si{\milli\tesla}$, yielding $\beta_e\simeq 0.4$ and $S\simeq 10^4$.

\section[sec:vpic]{2D and 3D Kinetic Simulations of TREX}

\begin{figure}
\includegraphics[width=\textwidth]{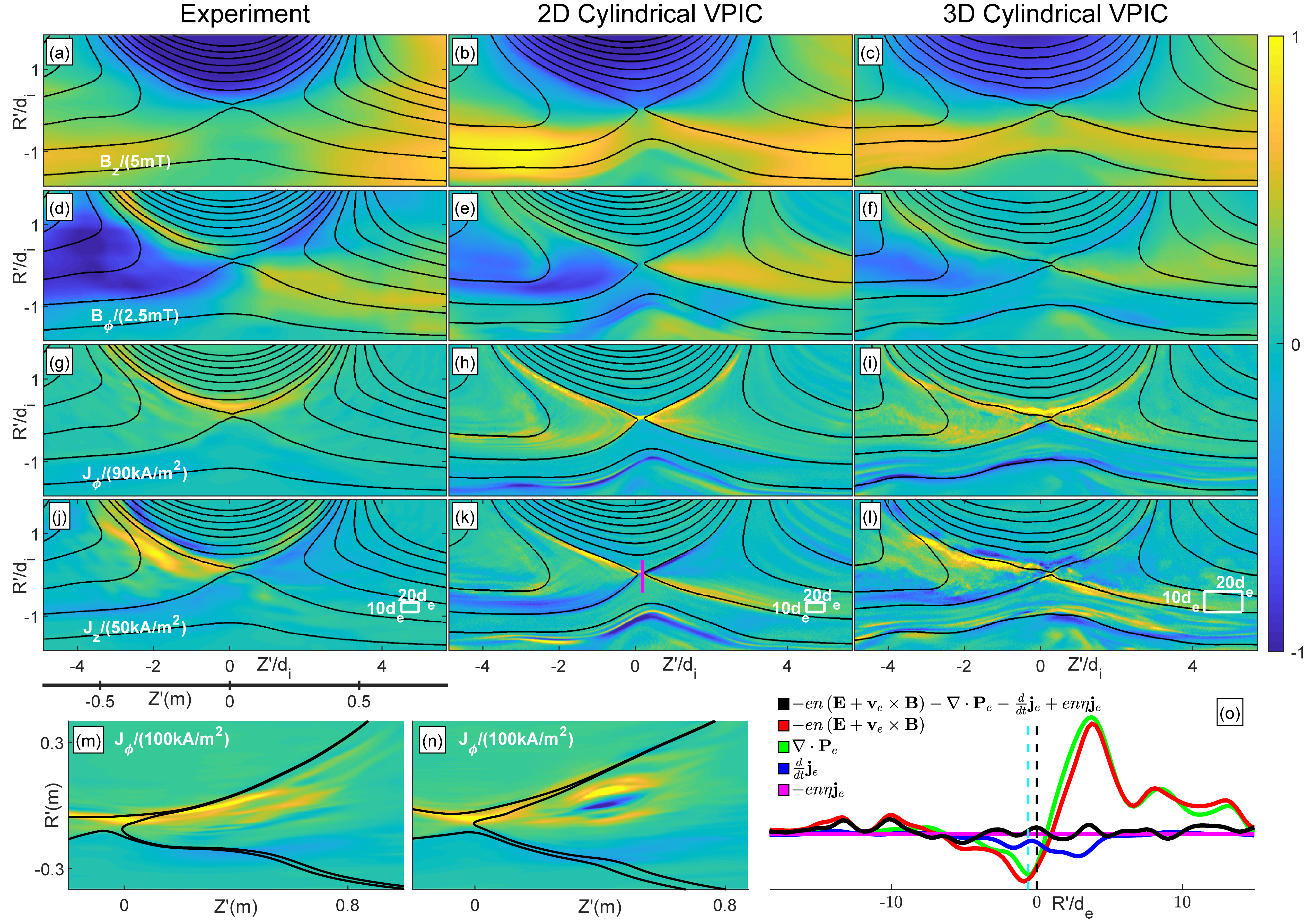}
\caption{Comparison of TREX experimental data from a scan of the hook probe (green in Fig.~\ref{fig:schematic}(b)) with cylindrical VPIC runs in both 2 and 3 dimensions. The 2D run is at full mass ratio, and the 3D run is at a mass ratio of 400. The 3D run plots show data taken from a single value of $\phi$. Each row of plots for a different magnetic feature shows contours of the magnetic flux function $\psi$ in black. Each row has been scaled relative to the same magnitude, shown in the leftmost plot of each row. (a-c) compare the in-plane ($B_z$); (d-f) show out-of-plane ($B_{\phi}$) magnetic fields. (g-i) show the out-of-plane current density ($J_{\phi}$); (j-l) show the in-plane ($J_z$) current structures. Plots (m) and (n) show experimental data from the linear probe (blue in Fig.~\ref{fig:schematic}(b)) using the jogging method. The jogging method provides a high spatial resolution of $\sim$ $0.4$ $\si{\centi\meter}$ in the $R$ direction. Plot (m) shows a mostly laminar layer with some bifurcation, whereas plot (n) shows a plasmoid being ejected from the reconnection region. Plot (o) is a standard Ohm's Law analysis of the 2D VPIC simulation, along the path indicated by the magenta line in plot (k). The dashed cyan line is the location of the $B_z$ $=$ $0$ point along the line; the dashed black line is the location of the maximum out-of-plane current density.} 
\label{fig:vpiccomp}
\end{figure}

TREX was simulated using VPIC, a kinetic particle-in-cell code \cite{bowers:2009,2018APS..DPPC11023D,bowers_2020}. The TREX boundary conditions were implemented in the new Cylindrical VPIC code with conducting walls at $R=1.5\si{\meter}$ and $Z= \pm 1.5\si{\meter}$, as well as an additional conducting wall at an adjustable minimum (nonzero) $R$ near the central axis. Within the simulation domain, current sources with the same dimensions as the TREX drive coils were added at the drive coil locations. The current density at these locations is increased as a function of time to mimic the ramping current injection utilized in the experiment. Using density data from TREX, initial density and magnetic field profiles were set at the simulation start time to balance the magnetic and kinetic pressures for a given applied Helmholtz field. Electron-electron, electron-ion, and ion-ion collisions were implemented in some 2D runs with a Monte-Carlo collision operator for binary Coulomb collisions \cite{takizuka1977}. The collision frequencies were calculated from TREX data. Not all simulations implemented collisions; results from testing a range of 2D simulation collision parameters at relevant experimental levels showed very little difference between runs with and without collisions. The 3D run discussed in this paper did not include collisions. 

The \change[]{grid spacing}{number of grid-points} in the 2D simulation described here was $1512$ by $3600$ in the $R$ and $Z$ directions, respectively; in 3D, these increased to $1024$ and $2048$ respectively with another $256$ grid divisions in the $\phi$ direction. The system size in 2D was about $193$ by $399$ electron skin depths in the $R$ and $Z$ directions, respectively. In 3D, these values were $87$ and $186$ electron skin depths, respectively. The average number of super-particles per cell was $500$ in 2D and $100$ in 3D. In both simulations, the ratio of the electron cyclotron frequency to the electron plasma frequency was $1$.

Both 2D $(RZ)$ and 3D $(R\phi Z)$ simulations of TREX can be compared to experimental results; one such comparison is shown in Fig.~\ref{fig:vpiccomp}. Experimental data in subplots (a,d,g,j) come from combining multiple shots worth of experimental data from the hook probe (green in Fig.~\ref{fig:schematic}(b)).

The 2D simulation (b,e,h,k) was obtained at full mass ratio while the 3D simulation (c,f,i,l) applies $m_i/m_e = 400$. For numerical tractability, the 3D simulation is limited to a $60^{\circ}$ wedge with periodic boundaries in $\phi$. Both the experimental and simulation profiles in Fig.~\ref{fig:vpiccomp}(a-l) are displayed with the domains normalized by the local ion skin depths. Here the local ion skin depth is obtained from the value of $n_e$ in the high-density inflow (e.g., the density value shown in Fig.~\ref{fig:exampledata}(c) at $R'\sim-0.1\si{\meter}$). The scaling of simulation variables relative to experimental ones was implemented using the technique described in \citeA{egedal:2019}, where temperature and magnetic field profile matching occurred near the X-line during the reconnection process.
Further similarities between TREX and 3D Cylindrical VPIC will be discussed later in this paper.

\section[sec:background]{Inferring the Reconnection rate in the TREX geometry}

In 3D geometries the rate of reconnection is not always trivial to define \cite{hesse1993}. However, given the nominal 2D experimental setup we can define the reconnection rate as the rate at which flux upstream of the reconnection region reconnects and moves downstream. 
Fig.~\ref{fig:ohm}(a) shows the different ``categories" of magnetic flux in the TREX cross-section. The red region contains field lines from the Helmholtz coil that go through the drive loop area and are upstream of the reconnection region. The blue lines are also upstream of the reconnection, but these represent the new magnetic flux injected into the system by the drive coils. Reconnection results in the downstream field lines, shown in green. The magenta field lines are those from the external Helmholtz coil that are initially above the internal drive coils and thus do not take part in the reconnection process.
We can then define the remaining unreconnected flux, $\Psi_B$, as the (red) magnetic flux between $R=0$ and the reconnection layer $\Psi_B = \int_0^{2\pi} \dd \phi \int_0^{R_{x}(\phi)} R B_z dR$, where the integral is taken at a constant $Z$ that matches the location of the $X$-line and from $R=0$ to $R=R_x$. Here $R_x(\phi)$ is the radius of the center of the current layer which is moving radially inwards and which in the simulation is observed to be a function of $\phi$. This path of this integral is represented by the cyan line in Fig.~\ref{fig:ohm}(a). 

Again, because $\Psi_B$ is the remaining un-reconnected magnetic flux, it is clear that $-d\Psi_B/dt$ is the rate at which magnetic flux is being reconnected, i.e. the reconnection rate.
There exists of course ambiguity in how to define the center of the reconnection layer $R_x(\phi)$, but as long as $R_x(\phi)$ correctly characterizes the inward motion of the reconnection layer it turns out that $-d\Psi_B/dt$ is largely unaffected by the differences between any reasonable choice of $R_x(\phi)$. By Faraday's law it is also clear that 
\begin{equation}
  \label{eq:Faraday}
  -\frac{d\Psi_B}{dt} = \oint_{R_x} (\mathbf{E} + \mathbf{v}_{R_x} \times \mathbf{B}) \cdot
  d\mathbf{l} \simeq \oint_{R_x} \mathbf{E} \cdot 
  d\mathbf{l} \quad.
\end{equation}
Note that in 3D for a particular choice of $R_x(\phi)$ the value of $B_z(R_x(\phi))$ could be finite and oscillate along $R_x(\phi)$, but it is reasonable to impose that for a valid choice of $R_x(\phi)$, the average value of $B_z(R_x(\phi))$ must be small. Meanwhile $ \mathbf{v}_{R_x}$ will be near constant and directed radially inward such that the average value of $ (\mathbf{v}_{R_x} \times \mathbf{B}) \cdot d\mathbf{l}$ also becomes small and can be neglected, as expressed in Eq.~\ref{eq:Faraday}.
  
In 3D configurations variations are permitted and likely present in $\mathbf{E}$ along 
$R_x(\phi)$. However, the local electric field may always be expressed on the form $\mathbf{E}= -\nabla \Phi - \partial \mathbf{A}/\partial t$, and because for any $\Phi$ we have 
$\oint_{R_x} \nabla \Phi \cdot d\mathbf{l} =0$, it becomes clear that the reconnection electric field defined as 
\begin{equation}
  \label{eq:Erec}
  E_{\rm rec} \equiv -\frac{1}{2\pi\left<R_x\right>} \frac{d\Psi_B}{dt} 
  = \frac{1}{2\pi\left<R_x\right>}
    \oint_{R_x} -\frac{\partial \mathbf{A}}{\partial t} \cdot 
  d\mathbf{l} \quad,
  \end{equation}
is a measure of the average toroidal inductive electric field, not directly dependent on any electrostatic electric fields $-\nabla \Phi$ which may be present in the reconnection region.

The physics that allows the electron fluid, with bulk velocity $\mathbf{v_e}$, to decouple from the motion of the magnetic field can be analyzed using the momentum equation of the electron fluid (the generalized Ohm's law), which takes the form:
\begin{eqnarray}
\mathbf{E}  =  - \mathbf{v_e} \times \mathbf{B} + 
\eta \mathbf{J_e} - \frac{1}{ne} \boldsymbol\nabla \cdot \mathbf{P_e} - \frac{m_e}{e}\frac{\dd \mathbf{v_e}}{\dd t}
\label{eq:ohme}
\end{eqnarray}
Here $\mathbf{P_e}$ is the electron pressure tensor, with elements $p_{ij} = m \int \left( u_i-v_{e,i}\right) \left( u_j-v_{e,j} \right) f \dd^3u$, $\mathbf{v_e}$ is the bulk electron fluid velocity, and $\dd/\dd t$ is the total convective derivative, $\dd/\dd t = \partial/\partial t + \mathbf{v_e} \cdot \boldsymbol\nabla$. 

In particular for the pressure tensor, we may split its contributions into its scalar and off-diagonal parts $\mathbf{P_e} = 
p_e\mathbf{I}+ \boldsymbol{\pi}$, where trace$(\boldsymbol{\pi})=0$. 
As discussed above, the reconnection rate $E_{\rm rec}$ is proportional to the $R_x(\phi)$-average of $\mathbf{E}$, and it becomes clear that while $-\nabla p_e$ contributions can be important to balance local electrostatic components of $\mathbf{E}$, the total contribution is $0$ because $\oint \nabla p_e \cdot d\mathbf{l} =0$. Thus, the $\boldsymbol\nabla \cdot \mathbf{P_e}$ term only contributes to reconnection through the off-diagonal stress in $\boldsymbol{\pi}$. 

An analysis of the generalized Ohm's Law for a 2D VPIC simulation of TREX is shown in Fig.~\ref{fig:vpiccomp}(o), where the terms of Ohm's Law are evaluated along the path defined by the magenta line in Fig.~\ref{fig:vpiccomp}(k). Given the boundary conditions of this simulation, this analysis does not include any form of spatial averaging, resulting in some fluctuation in the net Ohm's law term (the black line in Fig.~\ref{fig:vpiccomp}(o)); nonetheless, the pressure tensor divergence term (green) is clearly the dominant contributor to the reconnection electric field (red), consistent with prior 2D simulations of low-collisionality asymmetric reconnection, including \citeA{egedal:2018}. 

\section[sec:resultsohm]{Ohm's Law Results from 3D Simulations of TREX}

\begin{figure}
\includegraphics[width=\textwidth]{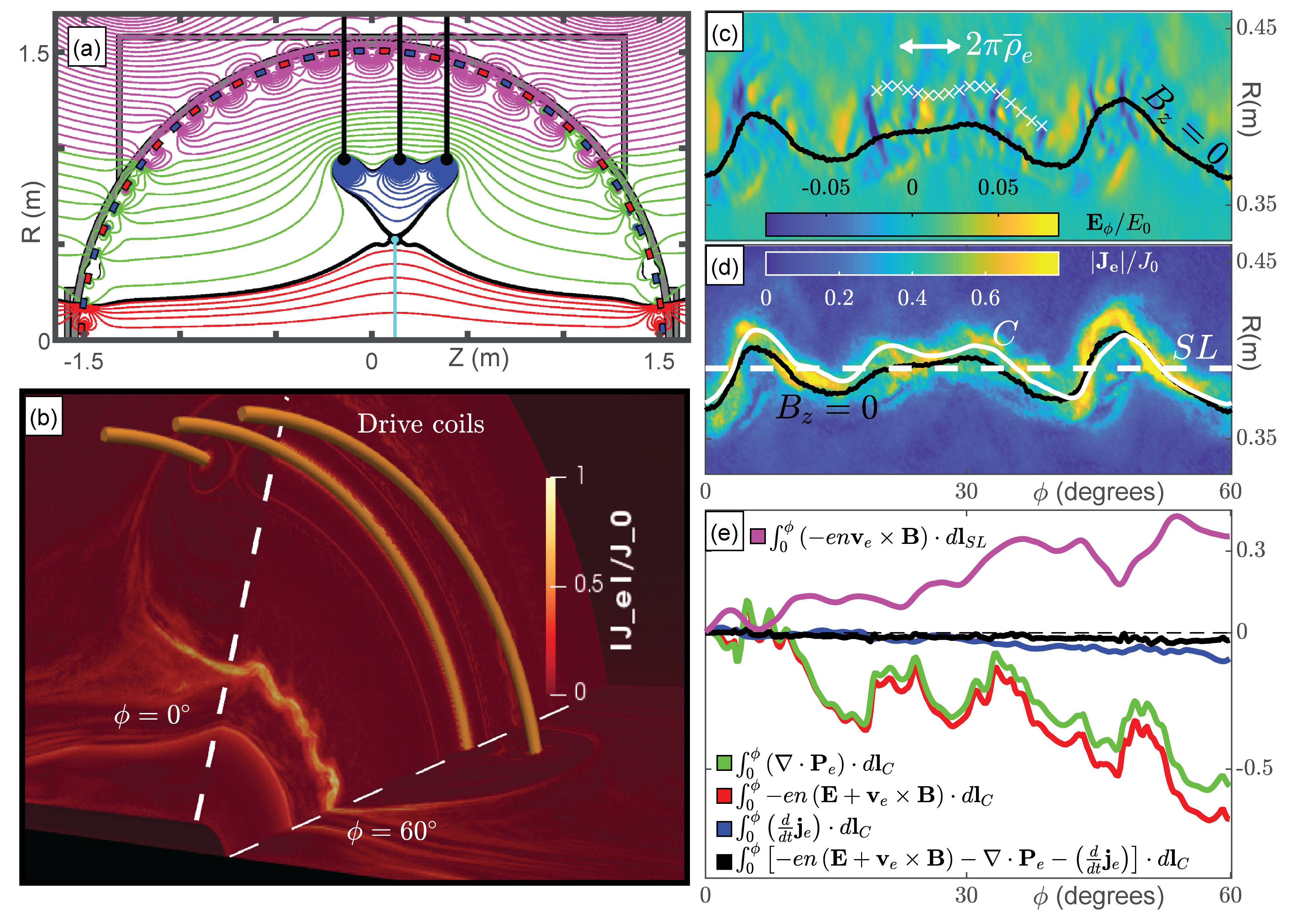}
\caption{ a) Block diagram of TREX's theoretical magnetic geometry, demonstrating the different regions of flux relative to the location of the reconnection layer. Blue and red are upstream of the reconnection region, starting from the drive coils and the Helmholtz coil, respectively. The green lines are downstream of the reconnection region, and the magenta lines are those that originate from the Helmholtz coil but are above the drive coils and thus do not impact the reconnection process. The cyan line represents the path of the integral used to define the flux function $\Psi$. b) Cuts of the current density, $|{\bf J}_e|$, in a 3D kinetic simulation of TREX implemented as a $60^{\circ}$ degree wedge with periodic boundaries in $\phi.$ The drive coils are shown as gold surfaces. (c) Simulation profiles of of ${\bf E}_\phi$ mapped onto the $R\phi$-plane at a single value of $Z$. The black line represents the $B_z$ $=$ $0$ line. Fluctuations in ${\bf E}_\phi$ have the characteristics of the electrostatic LHDI; notably, they are stronger above the layer where the particle density is lower. The average electron cyclotron radius, $\overline{\rho}_e$, is calculated along the path marked by the small white Xs; this is one of the primary scale lengths use to describe the two modes of the LHDI. (d) Profile of $|{\bf J}_e|$ mapped onto the $R\phi$-plane at a single value of $Z$; the black line is still the $B_z$ $=$ $0$ line. Also shown in white are two different paths of integration, a straight line (labeled SL) that simply cuts through the entire $\phi$ domain and the optimized integration curve $C$. The results of integrating the different terms of Ohm's Law along path $C$ are shown in (e); the Lorentz term (red) is almost completely matched by the pressure divergence term (green), and the net Ohm's Law term (black) is consistently negligible. Also displayed are the results of integrating solely the $\protect\mathbf{v_e}$ $\times$ $\protect\mathbf{B}$ term over the straight path $SL$ (shown in magenta).}
\label{fig:ohm}
\end{figure}

The 3D simulation domain can also be seen in Fig.~\ref{fig:ohm}(b). Taking a cut through the layer reveals the development of a toroidal instability which is inferred to be the lower hybrid drift instability (LHDI). Because the LHDI is driven by diamagnetic currents, it can be more vigorous within asymmetric reconnection layers which feature strong pressure gradients in the central portion \cite{roytershteyn:2012,le:2017,le:2018}, and several of the characteristics of the LHDI \cite{daughton:2003,davidson1977} matched the numerical layer fluctuations. The fastest growing LHDI modes are short wavelength ($\rho_e k_\perp \sim 2.9$), primarily electrostatic, and are localized on the edge of the layer; these can be seen in Fig.~\ref{fig:ohm}(c). However, the LHDI features a rich spectrum \cite{daughton:2003}, with longer wavelength $\sqrt{\rho_e \rho_i} k_{\perp} \sim 1.4$ electromagnetic modes that penetrate into the center, giving rise to a global rippling of the layer in the toroidal direction; this matches the characteristics of the kinking in the $B_z = 0$ and the current layer, as shown in Fig.~\ref{fig:ohm}(b)-(d).

Given the definition of the reconnection rate in Eq.~\ref{eq:Faraday}, we may now use the generalized Ohm's Law in Eq.~\ref{eq:ohme} to examine which term is responsible for making $ \oint_C {\bf E} \cdot d{\bf l}$ finite and thereby breaking the frozen-in condition for the electrons. While the value of $ \oint_C {\bf E} \cdot d{\bf l}$ is largely insensitive to the choice for $R_X(\phi)$, we will show that the magnitude of the terms on the RHS of Eq.~\ref{eq:ohme} greatly depend on this contour of integration; this has important implications for the amplitude of the aforementioned anomalous terms due to correlated spatial fluctuations. 

As mentioned, choosing a path over which to integrate is nontrivial; prior publications analysing Ohm's Law in similar parameter regimes have returned different results based on different choices of how to spatially average the relevant variables \cite{Price2017,le:2018}. A simple average over a single dimension (usually one analogous to what is here defined as our $\phi$ dimension) may pick up data from outside the diffusion region, leading to dominant terms that do not appear when an average is made over a path that has been adapted to fit the shape of the layer and any constituent instabilities \cite{che:2011,Price2017,le:2018}. 

Keeping this in mind, we chose to integrate over a layer-specific path while showing the potential consequence of selecting a simple single-variable path. The simple path is shown in Fig.~\ref{fig:ohm}(d) as a white dashed straight line labeled $SL$. The black path represents the contour along which $B_z = 0$; this path was used as a starting point for an iterative process that determined the layer-specific contour $C$ (solid white curve) by minimizing the contributions of the $-en\protect\mathbf{v_e} \times \protect\mathbf{B}$ term to the path integral.

Integrating Eq.~\ref{eq:ohme} along the path C produces the terms in Fig.~\ref{fig:ohm}(e). Note that collisionality was set to $0$ in this simulation, so the resistivity term of Eq.~\ref{eq:ohme} is also $0$. The left side of Eq.~\ref{eq:ohme} (the red line) is almost completely matched at every location by the pressure-tensor-divergence term (green line). The net Ohm's Law term (black) line remains near zero at every location. 

Also shown Fig.~\ref{fig:ohm}(e) is the $-en\protect\mathbf{v_e} \times \protect\mathbf{B}$ term for the $SL$ result; this is given by the magenta line. The kinking in the layer results in $SL$ including locations that are outside the dissipation region, resulting in large contributions of $-en\protect\mathbf{v_e} \times \protect\mathbf{B}$ which is the primary term for balancing $E_{\phi}$ outside the layer. By focusing on an average defined by $C$, we can minimize the $-en\protect\mathbf{v_e} \times \protect\mathbf{B}$ contributions and thus avoid drawing conclusions about the relevant terms of Ohm's Law in a manner that includes contributions from the non-reconnecting plasma regions. In previous analyses, contributions from the correlated fluctuations of the $-en\protect\mathbf{v_e} \times \protect\mathbf{B}$ term are combined with those from the pressure divergence to form the anomalous viscosity \cite{le:2018,che:2011}. These anomalous terms, when present, have been proposed as a mechanism that broaden the dissipation region and thus increase the reconnection rate \cite{parker:1957}; however, more recent results from reconnection dominated by kinetic effects show that this can be achieved without these anomalous terms becoming significant \cite{le:2018}. While there has been some disagreement about whether or not these anomalous terms are dominant in 3D kinetic reconnection \cite{che:2011,Price2017,le:2018}, our conclusion is that the answer will depend on the path used to define the spatial average. We find that the path that stays inside the diffusion region where $-en\protect\mathbf{v_e} \times \protect\mathbf{B}$ is small is the more physical choice. Furthermore, the dominant term breaking the electron frozen-in condition in the diffusion region is observed to be $\boldsymbol\nabla \cdot \mathbf{P_e}$ and is in agreement with recent spacecraft observations \cite{egedal:2019,nakamura2019}.

\section[sec:resultswidth]{Layer Width Results from Experiment and Simulation}

\begin{figure}
\includegraphics[width=\textwidth]{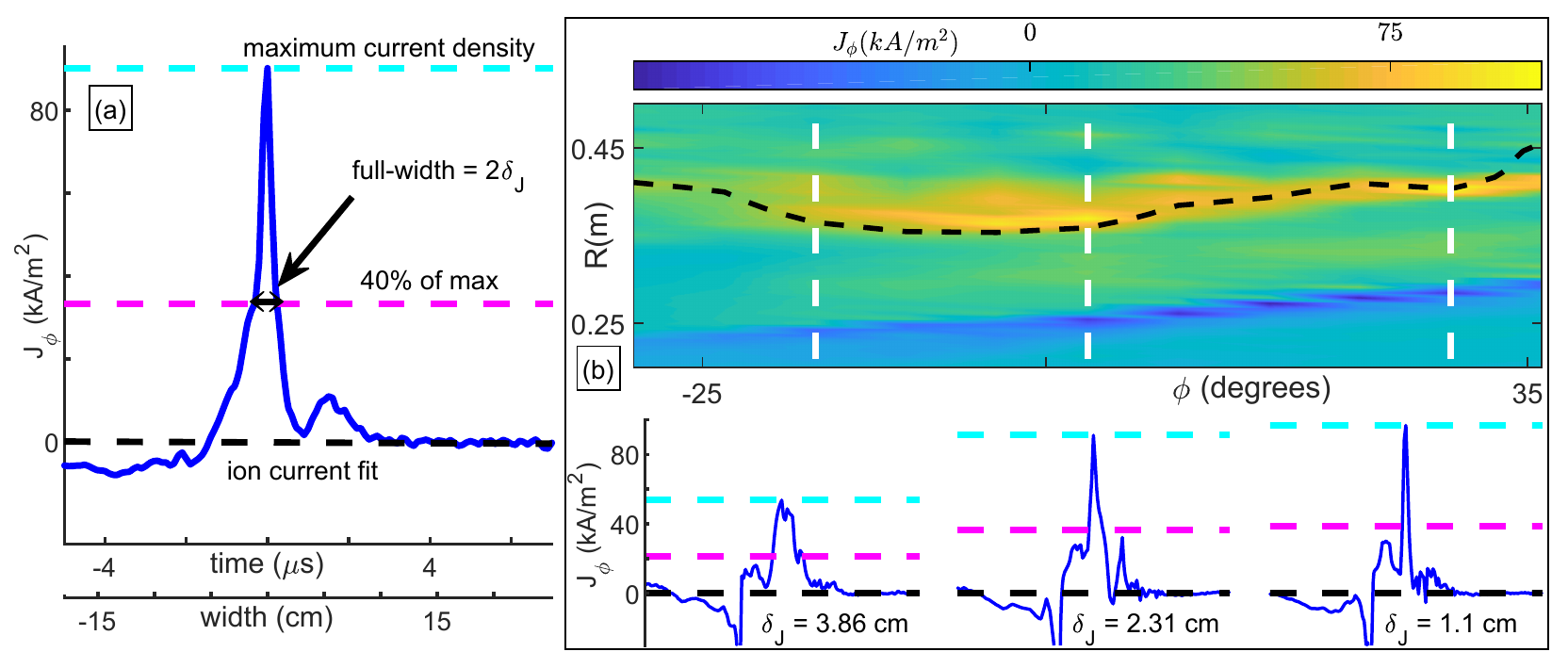}
\caption{Measuring layer width. (a) method for inferring the current layer half-width, $\delta_J$, defined as the half-width of the layer at $40\%$ of the maximum. (b) shows an $R\phi$ contour plot of the current layer in TREX recorded by a curved $\mathbf{\dot{B}}$ array sampling multiple toroidal angles. Three example slices at fixed $\phi$ are represented by the white dashed lines documenting variations in the layer widths.}
\label{fig:phiwidths}
\end{figure}

\begin{figure}
\includegraphics[width=\textwidth]{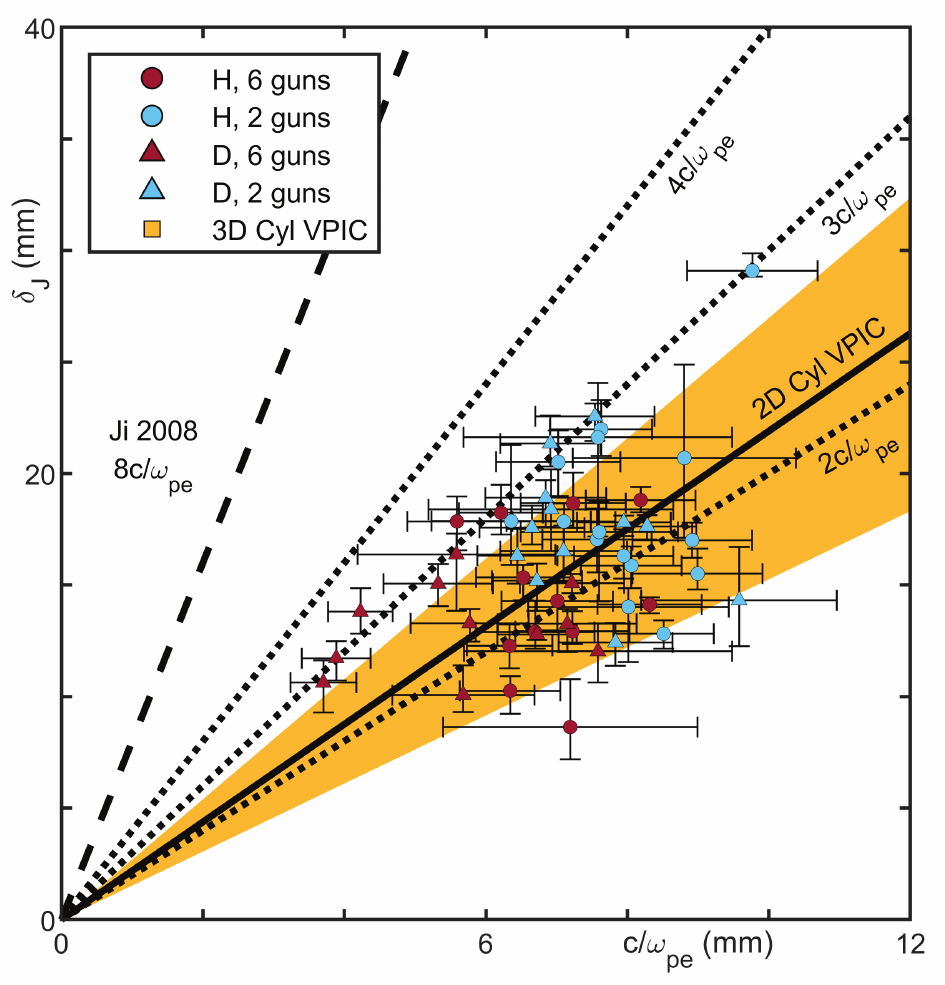}
\caption{Results of measuring experimental layer width over several different parameter sets. The colors highlight the difference between ``2 gun" (lower density) and a ``6 gun" (higher density) plasmas. The orange region shows the range of current layer widths measured in the 3D simulation, and the solid black line shows the result from a 2D axisymmetric simulation (without the instability). The spread in the experimental and 3D Cylindrical VPIC layer widths is attributed to the toroidal instability described in Fig.~\ref{fig:phiwidths}.}
\label{fig:layerwidth}
\end{figure}

In the above analysis of the kinetic simulation it was shown that the off-diagonal stress of $\boldsymbol\nabla \cdot \mathbf{P_e}$ is mainly responsible for breaking the electron frozen-in condition. From theory, this off-diagonal stress scales roughly as $1/\delta_J^2$, where $\delta_J$ is the half-width of the reconnection current layer \cite{hesse:1999}. Prior simulation results have established that when the current layer half-width $\delta_J$ is on the order of a few electron scale lengths (here taken to be the electron skin depth, $d_e=c/\omega_{pe}$, which at electron $\beta$ near unity is nearly equivalent to both the electron meandering width and the electron gyroradius), the pressure tensor divergence term will be sufficiently large to become the dominant term in Ohm's Law \cite{roytershteyn:2013}. This characteristic defines the regime of collisionless reconnection; as such, the measurement of layer widths on the order of the several $d_e$ can be used to establish the regime in which the reconnection is operating \cite{vasyliunas:1975,pritchett:2001}.

To address whether $\boldsymbol\nabla \cdot \mathbf{P_e}$ breaks the frozen-in condition in the experiment as it does in the simulation, we compare the widths of the current layers observed in TREX and the simulation. TREX is not yet able to measure the value of $\boldsymbol\nabla \cdot \mathbf{P_e}$ directly, but as described above, the existence of thin current layers implies that the $\boldsymbol\nabla \cdot \mathbf{P_e}$ contribution to Ohm's Law dominates the breaking of the frozen-in condition. The method of characterizing the layer half-width, $\delta_J$, is illustrated in Fig.~\ref{fig:phiwidths}(a), where the current density is recorded by the linear $\mathbf{\dot{B}}$ probe array (blue in Fig.~\ref{fig:schematic}(b)) at a set $\phi$ location for a range of $Z$ values. In addition, a toroidally curved $\mathbf{\dot{B}}$ probe array is used to characterize toroidal variations in the TREX current layers. An example dataset is shown in the top half of Fig.~\ref{fig:phiwidths}(b), documenting the intensity of the toroidal current density, with a range of different radial layer widths indicated below. This toroidal variation is similar to instability in the simulation, though the experimental observation is limited by the spatial resolution of the probe. 

Note that there is also some minor variation in the layer structure in the plane perpendicular to the toroidal direction; this can be seen in the existence of several smaller peaks in the current densities plotted in the bottom of Fig.~\ref{fig:phiwidths}(b). In this plane, the layer occasionally experiences some small amount of bifurcation or other minor irregularities. Examples of such behaviour are shown in Fig.~\ref{fig:vpiccomp}(m) and (n). In this analysis, only the primary peaks in the electron current density cuts are taken into consideration. 

Repeating the width measurement process for a range of experimental settings we obtain mean values for $\delta_J$ and $d_e$. The results are plotted in Fig.~\ref{fig:layerwidth}, where each data point represents the average result for $\delta_J$ and $d_e$ for a given set of drive potential, Helmholtz field, gun number, and ion species. There are approximately ten different experimental shots averaged for each data point, which include estimates for the experimental uncertainties. Also plotted is the line corresponding to the previously reported experimental results in MRX \cite{ji:2008}. The orange region represents the mean of the numerical layer widths from the 3D simulation in Fig.~\ref{fig:ohm}(b), $\pm$ a standard deviation. Similar to the experimental measurement process shown in Fig.~\ref{fig:phiwidths}(b), the numerical width results were obtained from radial cuts at different $\phi$ values relative to the phase of the instability. Finally, the width recorded in a corresponding 2D simulation in the $RZ$-plane is given by the solid black line. 

The width of the current layer is physically limited by the electron meandering scale, which is slightly larger than the electron inertial scale and has only a weak dependence on the precise $\beta_e$ \cite{dorfman:2008}; in these experiments this parameter is nearly constant ($\beta_e \sim 10^{-1}$). There is a clear divide between the blue (2 guns, lower density) and red (6 guns, higher density) datapoints, as expected from $d_e \propto n_e^{-1/2}$. Additionally, even though the lower density points have larger skin depths, they also have larger layer widths, keeping them on the same scaling as the higher density datapoints. There is not a particular relationship between ion species and the experimental width scaling. Most notably, there is a general spread in measured layer widths, both relative to different parameter sets and within a given parameter set itself (demonstrated by the vertical uncertainties). This is consistent with the presence of the toroidal instability measured in TREX and demonstrated in the 3D simulations. Crucially, both the absolute values and spread of the measured current layer widths is in good agreement between the simulation and the experiment.

\section[sec:conclusion]{Conclusions}

To summarize, reconnection in TREX is characterized by thin electron current layers, consistent with kinetic simulation results. The widths include a notable spread, $\delta_J \sim ( 1.5 - 3) d_e$, which can be attributed to the development of a toroidal instability in the current layer. Compared to previous experiments in MRX, the TREX temperature ratio $T_i/T_e\ll 1$ may be more favorable to this instability \cite{roytershteyn:2012}. Nevertheless, the TREX current layers are in good agreement with those observed in a 3D kinetic simulation, and are much thinner than those observed previously in the MRX experiment. 

The new Cylindrical VPIC code has allowed the TREX reconnection experiments to be modeled in a way that preserves its nominal cylindrical symmetry. Both 2D and 3D simulations reproduce the magnetic geometry measured in TREX, and 3D Cylindrical VPIC also shows the development of a toroidal instability that produces the same spread in the layer width scaling. The narrow current layers observed in TREX and their match to 3D kinetic simulation results validates the numerical result that off-diagonal stress in the electron pressure tensor is responsible for breaking the frozen-in condition for low collisionality configurations relevant to reconnection in the Earth's magnetosphere.

\acknowledgments
We gratefully acknowledge DOE funds DE-SC0019153, DE-SC0013032, and DE-SC0010463 and NASA fund 80NSSC18K1231 for support of the TREX experiment. In addition, the experimental work is supported through the WiPPL User Facility under DOE fund DE-SC0018266. Simulation work was supported by the DOE Basic Plasma Science program and by a fellowship from the Center for Space and Earth Science (CSES) at LANL. CSES is funded by LANL’s Laboratory Directed Research and Development (LDRD) program under project number 20180475DR. This work used resources provided by the Los Alamos National Laboratory Institutional Computing Program, which is supported by the U.S. Department of Energy National Nuclear Security Administration under Contract No. 89233218CNA000001. The 3D kinetic simulation was performed at the National Energy Research Scientific Computing Center (NERSC), a U.S. Department of Energy Office of Science User Facility operated under Contract No. DE-AC02-05CH11231. 

Data availability statement: Data from the 2D simulation is available at \citeA{greess_samuel_2021_4554697}. Data from the 3D VPIC simulation is available at \citeA{greess_samuel_2021_4556518}. Full figure data (both simulation and experimental) is available at \citeA{greess_samuel_2021_4837721}.

\bibliography{references}

\end{document}